\documentclass[twocolumn,
               showpacs,
               superscriptaddress,
               floatfix,
               longbibliography,
               aps,
               prl]{revtex4-2}

\usepackage[english]{babel}

\usepackage{graphicx}
\graphicspath{{./}}

\usepackage{amsmath,amssymb,bm,physics,upgreek,cancel,extarrows}
\usepackage[colorlinks=true,
            allcolors=blue]{hyperref}

\usepackage{acro}
\DeclareAcronym{bt}{
    short=BT,
    long=Bogoliubov transformation,
}
\DeclareAcronym{mgse}{
    short=MGSE,
    long=molecular ground state energy,
}
\DeclareAcronym{boa}{
    short=BOA,
    long=Born-Oppenheimer approximation,
}
\DeclareAcronym{bopes}{
    short=BOPES,
    long=Born-Oppenheimer potential energy surface,
    short-plural=s,
    long-plural=s,
}

\usepackage{comment}
\usepackage{subfiles}
\usepackage{lipsum}

\newcommand{\ed}{\downarrow}
\newcommand{\eu}{\uparrow}
\newcommand{\ii}{\mathrm{i}}
\newcommand{\ee}{\mathrm{e}}
\newcommand{\pii}{\uppi}

\newcommand{\figref}[2]{Fig.~\hyperref[#1]{\ref{#1}(#2)}}

\def\figArrays{
\begin{figure*}[t]%
    \centering%
    \includegraphics{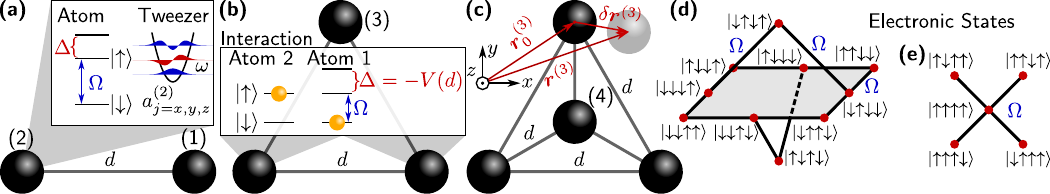}%
    \caption{\label{fig:arrays}%
             \textit{Atom tweezer arrays and electronic state space.} (a)
             Dumbbell-shaped atom array. Atoms are modeled by a two-level system
             with states $\ket*{\ed}$ (ground state) and $\ket*{\eu}$ (Rydberg
             state). Transitions are driven by a laser with detuning $\Delta$
             and Rabi frequency $\Omega$. The tweezer confinement is modelled by
             a harmonic oscillator with trapping frequency $\omega$ and phonon
             annihilation operators $a^{(k)}_j$ for the directions $j=x,y,z$.
             The superscript $k$ labels the atom in the array. The atomic
             equilibrium positions of the dumbbell are
             $\bm{r}^{(1)}_0=d(1,0,0)^\mathrm{T}$ and
             $\bm{r}^{(2)}_0=d(0,0,0)^\mathrm{T}$. (b) Triangular atom array,
             obtained by extending the dumbbell with a third atom at position
             $\bm{r}^{(3)}_0=d(1\slash{2},\sqrt{3}\slash{2},0)^\mathrm{T}$. If
             atom $2$ is in its Rydberg state, the Rydberg state of atom $1$
             experiences the interaction energy shift $V(d)$. When the laser
             detuning cancels this interaction shift, $\Delta=-V(d)$, atom $1$
             undergoes a facilitated excitation. (c) Tetrahedral atom array,
             obtained from the triangle [see (b)] by adding an atom at position
             $\bm{r}^{(4)}_0=d(1\slash{2},1\slash(2\sqrt{3}),\sqrt{2\slash{3}})
             ^\mathrm{T}$. The position
             $\bm{r}^{(k)}=\bm{r}^{(k)}_0+\delta\bm{r}^{(k)}$ of the $k$-th atom
             in the array is the sum of the equilibrium position
             $\bm{r}^{(k)}_0$ and the displacement vector $\delta\bm{r}^{(k)}$,
             with components
             $\delta{r}^{(k)}_j=x_0(a^{(k)}_j+(a^{(k)}_j)^\dagger)\slash
             \sqrt{2}$. (d,e) Visualization of coupled (at rate $\Omega$)
             resonant electronic states of the tetrahedron. The graph in (d)
             emerges when choosing the facilitation condition $\Delta=-V(d)$,
             and (e) is obtained when setting $\Delta=-3V(d)$.
            }%
\end{figure*}%
}
\def\figInstabilities{
\begin{figure}[t]%
    \centering%
    \includegraphics{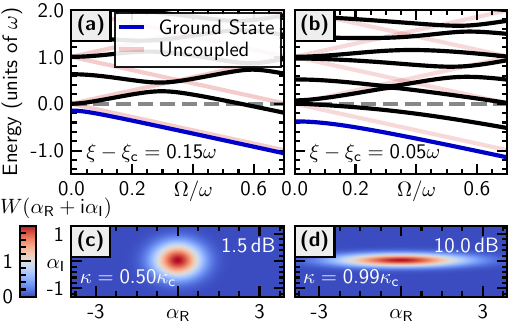}%
    \caption{\label{fig:instabilities}%
             \textit{Energy spectrum and Wigner distribution.} (a,b) Energy
             spectrum (black) of the dumbbell [\figref{fig:arrays}{a}] as a
             function of the Rabi frequency $\Omega$. The spectrum is computed
             by exact diagonalization of Eq.~\eqref{eq:dumbbellHamiltonian} with
             $\kappa=0.1\omega$ and truncating at 100 phonons. In blue, we
             highlight the ground state. As a reference we show the uncoupled
             ($\kappa=\xi=0$) spectrum (red). (c,d) Wigner quasiprobability
             distribution of the ground state of the tetrahedron
             [\figref{fig:arrays}{c}], Eq.~\eqref{eq:wignerDistribution}, for
             different values of $\kappa$. We choose $\xi=2\omega$ and
             $\nu=0.1$. In the top right the squeezing strengths (variance of
             the position quadrature associated to the mode with annihilation
             operator $b^{\perp,1}_{\ket*{\eu\eu\ed\ed}}$ normalized to the
             variance for $\kappa=\xi=0$) are shown.
             }%
\end{figure}%
}
\def\figBoa{
\begin{figure}[t]%
    \centering%
    \includegraphics{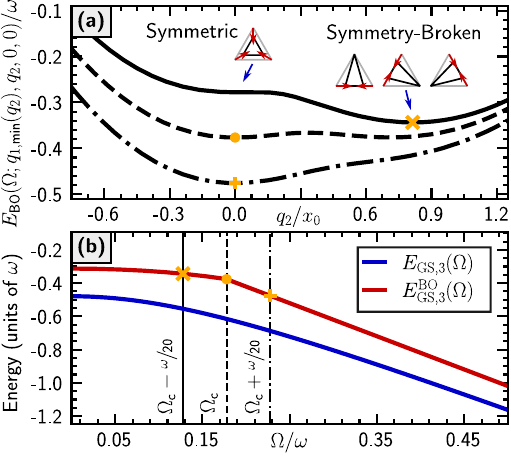}%
    \caption{\label{fig:boa}%
             \textit{Potential energy surfaces and ground state energy.} (a)
             Ground state potential energy surface for the triangle shown in
             \figref{fig:arrays}{b}. We plot the surface in the vicinity of one
             of the three-fold degenerate minima; the one that is associated
             with configurations in the vicinity of the leftmost of the three
             deformed triangular shapes. For $\Omega=0$ this surface is given by
             Eq.~\eqref{eq:potentialEnergySurface}. The curves shown here are
             $E_\mathrm{BO}(\Omega;q_{1,\mathrm{min}}(q_2),q_2,q_3=0,q_4=0)$,
             and the minimum of each curve (see yellow markers) corresponds to
             the ground state within the \ac{boa} for a given $\Omega$. At
             $\Omega=\Omega_\mathrm{c}$ a structural transition between a
             symmetric and a symmetry-broken triangle takes place (dashed line).
             For details see the supplemental material \cite{supmat}. For the
             electron-phonon coupling parameters we chose $\xi=-0.15\omega$,
             $\kappa=0.25\omega$, and $\nu=0.1$. (b) Ground state energy of the
             triangle as a function of the Rabi frequency $\Omega$. The curve
             $E_{\mathrm{GS},3}$ (blue) is obtained by exact diagonalization
             (truncated at 20 phonons per mode). The red curve shows the ground
             state energy $E^\mathrm{BO}_{\mathrm{GS},3}$, obtained in the
             \ac{boa}.
            }%
\end{figure}%
}

\begin{document}

\title{Rydberg atom arrays as quantum simulators for molecular dynamics}

\author{Simon Euchner}
\affiliation{Institut f\"ur Theoretische Physik and Center for Integrated
             Quantum Science and Technology, Universit\"at Tübingen,
             Auf der Morgenstelle 14, 72076 T\"ubingen, Germany}
\author{Igor Lesanovsky}
\affiliation{Institut f\"ur Theoretische Physik and Center for Integrated
             Quantum Science and Technology, Universit\"at Tübingen,
             Auf der Morgenstelle 14, 72076 T\"ubingen, Germany}
\affiliation{School of Physics and Astronomy and Centre for the Mathematics and
             Theoretical Physics of Quantum Non-Equilibrium Systems, The
             University of Nottingham, Nottingham, NG7 2RD, United Kingdom}

\begin{abstract}
    Rydberg atoms held in optical tweezer arrays combine vibrational and
    electronic degrees of freedom which can be coupled and manipulated at a
    microscopic level. This opens opportunities for the quantum simulation of
    artificial molecular systems and offers in particular a platform for probing
    complex vibronic dynamics in controlled settings with increasing complexity.
    Tailored interatomic interactions and electron-phonon couplings yield
    handles for designing electronic state manifolds, for studying structural
    transitions, and for exploring non-classical vibrational states near
    molecular instabilities. Furthermore, this quantum simulator opens
    opportunities for testing and quantifying the validity of fundamental
    concepts, such as the Born-Oppenheimer approximation and quantum corrections
    to it.
\end{abstract}

\maketitle

\textit{Introduction.---}Shedding light on dynamical processes within molecules
is essential for many problems in physics, chemistry, and biology
\cite{pauling1931,lou2024,li2024,goossens2018}. However, their coupled
electronic and motional degrees of freedoms typically result in a
high-dimensional state space that makes \emph{ab initio} calculations
forbiddingly complex \cite{atkins2011}. This is particularly the case when
common simplifications, such as the adiabatic approximation \cite{born1927}, for
computing electronic potential surfaces are not applicable. An important example
in this regard is the process of human vision which relies on a conformational
change of molecular structure near crossing potential surfaces
\cite{shoenlein1991,rinaldi2014}. This and other related processes take place
usually on an Angstrom length scale and on a femtosecond time scale. One
approach to study them at directly observable length and drastically prolonged
time scales --- and thereby to shed light on their physics --- is to use quantum
simulators \cite{manin1980,feynman1982}. Initial steps into this direction have
already been undertaken: trapped ion setups have been used to mimic molecular
dynamics near so-called conical intersections
\cite{gambetta2021,whitlow2023,valahu2023}. Rydberg atoms have been employed to
create macroscopically large Rydberg molecules
\cite{boisseau2002,overstreet2009,bendkowsky2009,kiffner2012,sassmanhausen2016,
shaffer2018,exner2025} as well as aggregates and clusters
\cite{wuester2018,schempp2014}. The latter were probed using the van der Waals
explosion \cite{faoro2016}, in analogy to the Coulomb explosion
\cite{vager1989}, which is a means to infer molecular structure.
\par
In this paper, we showcase the potential of neutral atom arrays in optical
tweezers \cite{schymik2020,endres2016,barredo2016,barredo2018,schlosser2023} to
serve as a quantum simulation platform for the study of molecular processes. To
illustrate the capabilities of the simulator we focus on three elementary
molecular shapes of increasing complexity --- a dumbbell, a triangle, and a
tetrahedron. Already these simple examples reveal a range of intriguing
properties, such as the emergence of motional instabilities and non-classical,
i.e., squeezed, vibrational ground states. The platform furthermore offers the
capability of designing the structure of the electronic state space. More
broadly, it allows to systematically investigate the impact of quantum effects,
which become particularly important in regimes where the commonly used \ac{boa}
\cite{born1927} breaks down. Rydberg tweezer arrays allow to quantitatively test
the applicability of such approximations and to perform fully quantum mechanical
investigations of molecular processes in system sizes that are intractable by
exact numerical methods.

\figArrays 
\textit{Molecular quantum simulator.---}We consider $N$ optical tweezers, each
confining a single atom [see examples in \figref{fig:arrays}{a-c}]. The dynamics
of a single atom trapped in the $k$-th tweezer is described by the single atom
Hamiltonian ($\hbar\equiv{1}$)
\begin{equation}\label{eq:singleAtomHamiltonian}
    H^{(k)} = \Omega \sigma_x^{(k)}
              +
              \Delta n^{(k)}
              +
              \omega \sum_{j=x,y,z} (a^{(k)}_j)^\dagger a^{(k)}_j
\,.
\end{equation}%
The first two terms, with $\sigma_x=\dyad*{\ed}{\eu}+\dyad{\eu}{\ed}$ and
$n=\dyad{\eu}{\eu}$, model a laser which drives transitions between the ground
($\ket*{\ed}$) and Rydberg ($\ket*{\eu}$) state with Rabi frequency $\Omega$ and
detuning $\Delta$ [cf. \figref{fig:arrays}{a}]. The second term describes the
vibrations of the atom in the tweezer relative to its equilibrium position.
Here, the phonon annihilation operators, $a^{(k)}_j$, represent the
corresponding displacements
$\delta{r}^{(k)}_j=x_0\big(a^{(k)}_j+(a^{(k)}_j)^\dagger\big)\slash\sqrt{2}$ in
direction $j=x,y,z$ [see \figref{fig:arrays}{c}], and
$x_0=1\slash\sqrt{m\omega}$ is the harmonic oscillator length. For simplicity,
we assume the tweezer potentials to be isotropic and electronic
state-independent \cite{ahlheit2024,zhang2011}. The full many-body Hamiltonian
of the tweezer array is
\begin{equation}\label{eq:simulatorHamiltonian}
    H =   \sum_{k=1}^N H^{(k)}
        + \sum_{k=1}^N \sum^{k-1}_{l=1}
          V(r^{(k,l)}) n^{(k)} n^{(l)}
\,,
\end{equation}%
where $V(r^{(k,l)})$ describes interactions between Rydberg atoms as a function
of the distance $r^{(k,l)}$ between the $k$-th and $l$-th atom. Typically, the
interaction potential follows a power-law
\cite{saffman2010,sibalic2017,singer2005}. However, using F\"orster resonances
and microwave dressing allows one to create a variety of shapes
\cite{parismandoik2016,petrosyan2014,foerster1948}, including Lennard-Jones type
potentials with repulsive core and attractive long-range tail. Typical atom
equilibrium distances in tweezer arrays are
$d\approx{3\text{--}6\,\text{\textmu{m}}}$
\cite{labuhn2014,browaeys2020,nogrette2014} and trap frequencies can be on the
order of $\omega=100\,\mathrm{kHz}\approx{2}\pii\cdot{16}\,\mathrm{kHz}$. For
$^7\mathrm{Li}$ atoms this yields a harmonic oscillator length of
$x_0=1\slash\sqrt{\omega{m}}\approx{300}\,\mathrm{nm}$ and hence a ratio between
atomic displacement and interatomic separation of
$\nu=x_0\slash{d}\lesssim{0.1}$. Close to the vibrational ground state this
justifies the following expansion of the interaction potential
\begin{multline}\label{eq:potentialExpansion}
    V(r^{(k,l)})
    \approx
        V(r^{(k,l)}_0)
      +
        \bm{G}^{(k,l)}\delta\bm{r}^{(k,l)}
\\
      +
        \frac{1}{2}
        (\delta\bm{r}^{(k,l)})^\mathrm{T}
        \big(H^{(k,l)}_{\mathrm{a}}+H^{(k,l)}_{\mathrm{b}}\big)
        \delta\bm{r}^{(k,l)}
\,,
\end{multline}%
with $\delta\bm{r}^{(k,l)}=\delta\bm{r}^{(k)}-\delta\bm{r}^{(l)}$. Here the
second and third term couple the electronic state-dependent interaction to the
vibrations of the atoms inside the tweezer potential. The strength and nature of
this vibronic coupling is quantified by the expansion coefficients
\begin{align}
    \begin{split}\label{eq:gradientStructure}
        \bm{G}^{(k,l)}
        &=
        V'(r_0^{(k,l)})\frac{(\bm{r}^{(k,l)}_0)^\mathrm{T}}{r^{(k,l)}_0}
    \end{split}
\,,\\
    \begin{split}\label{eq:hessianAStructure}
        H^{(k,l)}_{\mathrm{a}}
        &=
        V''(r^{(k,l)}_0)
        \frac{\bm{r}^{(k,l)}_0}{r^{(k,l)}_0}
        \otimes
        \frac{(\bm{r}^{(k,l)}_0)^\mathrm{T}}{r^{(k,l)}_0}
    \end{split}
\,,\\
    \begin{split}\label{eq:hessianBStructure}
        H^{(k,l)}_{\mathrm{b}}
        &=
        \frac{V'(r^{(k,l)}_0)}{r^{(k,l)}_0}
        \bigg[
            1_3
            -
            \frac{\bm{r}^{(k,l)}_0}{r^{(k,l)}_0}
            \otimes
            \frac{(\bm{r}^{(k,l)}_0)^\mathrm{T}}{r^{(k,l)}_0}
        \bigg]
    \end{split}
\,,
\end{align}%
where $\bm{r}^{(k,l)}_0=\bm{r}^{(k)}_0-\bm{r}^{(l)}_0$ denotes the distance
vector between equilibrium positions and $r^{(k,l)}_0=\norm*{\bm{r}^{(k,l)}_0}$.

\textit{Molecular Hamiltonians.---}To get a first impression of the capabilities
of the simulator we first focus on the dumbbell depicted in
\figref{fig:arrays}{a}. In order to simplify the problem we employ the
facilitation condition $\Delta=-V(d)$ and $\abs*{\Delta}\gg\abs*{\Omega}$
\cite{su2020,pohl2007,jau2016,kurdak2024, ates2007,marcuzzi2017} [see
\figref{fig:arrays}{b}]. This splits the electronic Hilbert space into two
approximately decoupled \cite{james2007} manifolds of resonant states:
$\{\ket*{\ed\ed}\}$ and $\{\ket*{\ed\eu},\ket*{\eu\ed},\ket*{\eu\eu}\}$, with
energy separation $\abs*{\Delta}$. Only the latter one features vibronic
couplings. Here, the constrained \cite{mazza2020,magoni2021,brady2025}
electronic dynamics can be represented by a tight-binding model on a graph,
where $\Omega$ is the coherent hopping rate between nodes:
$\ket*{\ed\eu}\xleftrightarrow{\Omega}\ket*{\eu\eu}
\xleftrightarrow{\Omega}\ket*{\eu\ed}$. In fact a further simplification is
achieved by introducing the state
$\ket*{+}=(\ket*{\ed\eu}+\ket*{\eu\ed})\slash\sqrt{2}$, which restricts the
electronic dynamics to the space $\{\ket*{+},\ket*{\eu\eu}\}$. Furthermore,
introducing the vibrational mode of the relative motion,
$b=(a^{(1)}-a^{(2)})\slash\sqrt{2}$, yields the following compact matrix
representation of the dumbbell Hamiltonian:
\begin{equation}\label{eq:dumbbellHamiltonian}
    H_2 = \mqty[ \omega b^\dagger b & \sqrt{2}\Omega
                 \\
                 \sqrt{2}\Omega &   \omega b^\dagger b
                                  + \sqrt{2}\kappa(b + b^\dagger)
                                  + \xi (b+b^\dagger)^2           ]
\,.
\end{equation}%
Here, we defined the vibronic coupling constants $\kappa=x_0V'(d)\slash\sqrt{2}$
and $\xi=x_0^2V''(d)\slash{2}$ which depend on the interaction potential
gradient and curvature at the interatomic equilibrium distance $d$. This
Hamiltonian illustrates the general idea behind the quantum simulator of
molecular phenomena: diagonal entries describe the motion of ``nuclear" degrees
of freedom in electronic potential energy surfaces, which are a combination of
the tweezer trap and the state-dependent interaction potential. These surfaces
are hybridized by the off-diagonal entries due to the laser coupling with Rabi
frequency $\Omega$. In general, the shape of the Hamiltonian can thus be
engineered by controlling the Rydberg interaction potential $V$, i.e., through
$\kappa$ and $\xi$, the structure of the electronic state space, and the
geometric arrangement of the tweezers. To illustrate this further, we consider
next the tetrahedron depicted in \figref{fig:arrays}{c}. More atoms yield a
larger electronic Hilbert space and choosing the laser detuning $\Delta$ allows
to engineer different graphs of resonant electronic states: setting
$\Delta=-V(d)$ yields the multiple-connected graph in \figref{fig:arrays}{d} and
setting $\Delta=-3V(d)$ leads to the star graph shown in \figref{fig:arrays}{e}.
Generally, the emerging vibronic Hamiltonians take the form
\begin{equation}\label{eq:tetrahedronHamiltonian}
    H_\textrm{mol}
    =
        \Omega \sum_{s s'} A_{s s'} \dyad*{s}{s'}
      +
        \sum_{s} h_s \dyad*{s}{s}
\,,
\end{equation}%
where the matrix $A$ is the adjacency matrix of the graph representing the
electronic states that are resonantly connected by the laser. Furthermore, the
Hamiltonians $h_s$ model the electronic state-dependent vibrational ``nuclear"
dynamics. For the particular case of the tetrahedron and the graph in
\figref{fig:arrays}{d} $A$ is ten-dimensional and the indices $s$ and $s'$ run
over all electronic states contained in the graph. In states containing one
Rydberg excitation vibrations are solely governed by the tweezer confinement:
$h_s=\omega\bm{b}^\dagger_s\bm{b}_s$, where we collected all modes in the vector
$\bm{b}_s$. States containing two Rydberg atoms feature a significantly more
complex vibrational dynamics, which is governed by 
\begin{multline}\label{eq:bosonicHamiltonians}
    h_s
    =
        \omega \bm{b}^\dagger_s \bm{b}_s
      +
        \sqrt{2}\kappa \big(b^\parallel_s+(b^\parallel_s)^\dagger\big)
      +
        \xi\big(b^\parallel_s+(b^\parallel_s)^\dagger\big)^2
\\
      +
        \frac{\nu\kappa}{\sqrt{2}}
        \big(b^{\perp,1}_s+(b^{\perp,1}_s)^\dagger\big)^2
      +
        \frac{\nu\kappa}{\sqrt{2}}
        \big(b^{\perp,2}_s+(b^{\perp,2}_s)^\dagger\big)^2
\,.
\end{multline}%
The parallel modes describe displacements along the tetrahedron's edges,
analogously to $b$ in Eq.~\eqref{eq:dumbbellHamiltonian}. The perpendicular
modes govern motion within the plane perpendicular to each edge. For details see
the supplemental material \cite{supmat}.

\figInstabilities 
\textit{Molecular instabilities and ground state squeezing.---}Let us now
discuss some phenomena which can be explored with the quantum simulator. To this
end we return to the dumbbell in Eq.~\eqref{eq:dumbbellHamiltonian} and consider
the limit $\Omega \rightarrow 0$. Here the Hamiltonian can be diagonalized by a
state-dependent \ac{bt}. In the $\ket*{+}$-state this is trivial, whereas in the
$\ket*{\eu\eu}$-state we have the transformed mode (see supplemental material
\cite{supmat}) $b_\mathrm{BT}=(b+wb^\dagger)\slash\sqrt{1-w^2}$, where
\begin{equation}\label{eq:solutionForW}
    w
    =
    1+\phi-\frac{\phi}{\abs*{\phi}}\sqrt{(1+\phi)^2-1}
\,, \
    \phi = \frac{\omega}{2\xi}
\,.
\end{equation}%
Crucially, $w$ exists only if $\xi>\xi_\mathrm{c}$, where
$\xi_\mathrm{c}=-\omega\slash{4}$ is the critical potential curvature. Outside
this regime the electron-phonon coupling renders the system unstable. This
becomes also apparent from the \ac{mgse}, which for $\Omega=0$ reads
$E_{\mathrm{GS},2}=\mathrm{min}\{\omega\varepsilon_2,0\}$, with
\begin{equation}\label{eq:groundStateEnergyDumbbell}
    \varepsilon_2
    =
      - \frac{2\kappa^2}{\omega^2} \frac{1}{1-\bar{\xi}}
      + \frac{1}{2}\sqrt{1-\bar{\xi}}
      - \frac{1}{2}
\,, \
    \bar{\xi} = \frac{\xi}{\xi_\mathrm{c}}
\,.
\end{equation}%
At $\xi=\xi_\mathrm{c}$ the denominator of the first term vanishes and the
energy diverges. For finite $\Omega$, we computed the \ac{mgse} numerically,
which is shown as the blue curve in \figref{fig:instabilities}{a,b}. Clearly,
the ground state energy is lowered as $\abs*{\xi-\xi_\mathrm{c}}$ decreases and
achieving numerical convergence becomes increasingly difficult as the
instability is approached. In fact, the instability at $\xi_\mathrm{c}$ is
expected to occur for arbitrary Rabi frequencies $\Omega$ because
Eq.~\eqref{eq:solutionForW} does not depend on $\Omega$. Physically, we
understand the instability at $\xi_\mathrm{c}$ as the consequence of the
trapping potential being canceled by the quadratic perturbations of the harmonic
oscillator in Eq.~\eqref{eq:dumbbellHamiltonian}. This is a somewhat trivial
effect. The tetrahedron on the other hand features a second, in fact more
interesting, instability which we find upon inspection of its ground state
energy (at $\Omega=0$). It is given by
$E_{\mathrm{GS},4}=\mathrm{min}\{\omega\varepsilon_4,0\}$, with
\begin{equation}\label{eq:groundStateEnergyTetrahedron}
    \varepsilon_4
    =
      - \frac{2\kappa^2}{\omega^2}\frac{1}{1-\bar{\xi}}
      + \frac{1}{2}\sqrt{1-\bar{\xi}}
      + \sqrt{1-\bar{\kappa}}
      - \frac{3}{2}
\,, \
    \bar{\kappa} = \frac{\kappa}{\kappa_\mathrm{c}}
\,.
\end{equation}%
This expression is obtained by employing the \ac{bt} defined by
Eq.~\eqref{eq:solutionForW} to the Hamiltonians $h_s$ in the state space
contained in the graph of \figref{fig:arrays}{d}. A direct inspection shows
that, in addition to the instability at $\xi_\mathrm{c}$, we obtain a second one
at $\kappa_\mathrm{c}=-\omega\slash(2\sqrt{2}\nu)$, manifesting in an imaginary
square root when $\kappa<\kappa_\mathrm{c}$. This second instability can be
characterized through the squeezing of the ground state wave function: for
$\Omega=0$ a ground state of Eq.~\eqref{eq:tetrahedronHamiltonian} is the
product $\ket*{\eu\eu\ed\ed}\otimes\ket*{\mathrm{GS}}$, where
$\ket*{\mathrm{GS}}$ is the ground state of $h_{\ket*{\eu\eu\ed\ed}}$ defined in
Eq.~\eqref{eq:bosonicHamiltonians}. Its Wigner quasiprobability distribution for
the mode $b^{\perp,1}_{\ket*{\eu\eu\ed\ed}}$ is, with
$\alpha=\alpha_\mathrm{R}+\ii\alpha_\mathrm{I}$,
\begin{equation}\label{eq:wignerDistribution}
    W(\alpha)
    =
    \frac{2}{\pii}
    \exp(-\bar{w}_+\alpha_\mathrm{R}^2-\bar{w}_-\alpha_\mathrm{I}^2)
    \,, \
    \bar{w}_\pm = 2\frac{1\pm w}{1\mp w}
    \,,
\end{equation}%
as derived in the supplemental material \cite{supmat}. We plot this function for
different values of $\kappa$ in \figref{fig:instabilities}{c,d}, to show the
increasingly squeezed shape for $\kappa\searrow\kappa_\mathrm{c}$ when
approaching the instability. Importantly, although we are close to the critical
point, $\kappa=0.99\kappa_\mathrm{c}$, where one could expect large ground state
displacements because of weak atom confinement, the atoms are in fact on average
closer than $0.4x_0$ to their equilibrium positions (see supplemental material
\cite{supmat}). We note here, that going beyond the considered limit,
$\Omega={0}$, requires exact numerical diagonalization. This is, however, rather
challenging, given that already for such a small system --- with all symmetries
considered --- $9$ vibrational modes and $10$ electronic states [see
\figref{fig:arrays}{d}] are involved.

\figBoa
\textit{Structural molecular transitions.---}Finally, let us show that this
quantum simulation platform permits the investigation of structural molecular
transitions and allows to study the impact of quantum effects on those. To this
end we consider the triangle shown in \figref{fig:arrays}{b}, which can undergo
a transition from a symmetric to a symmetry-broken ground state. Employing the
facilitation condition, $\Delta=-V(d)$, its resonant electronic subspace is
spanned by the states $\{\ket*{\ed\ed\eu}$, $\ket*{\ed\eu\ed}$,
$\ket*{\ed\eu\eu}$, $\ket*{\eu\ed\ed}$, $\ket*{\eu\ed\eu}$,
$\ket*{\eu\eu\ed}\}$. The resulting Hamiltonian has the structure of
Eq.~\eqref{eq:tetrahedronHamiltonian} with $b^{\perp,2}_s=0$ and an adjacency
matrix that describes a ring-shaped graph \cite{magoni2023}. To see the
structural transition it is sufficient to employ the so-called \ac{boa}, which
amounts to neglecting the kinetic energy of the ``nuclei". The resulting
\acp{bopes} depend on four collective coordinates (bosonic modes), which we
label $q_i$, $i=1,2,3,4$ (see supplemental material \cite{supmat}). To compute
the \ac{mgse} we minimize the (classical) \ac{bopes}
$E_\mathrm{BO}(\Omega;\bm{q})$, defined as the smallest eigenvalue of the
Hamiltonian for a fixed position $\bm{q}=(q_1,q_2,q_3,q_4)^\mathrm{T}$ and Rabi
frequency $\Omega$. Note, that for $\Omega=0$ there are three degenerate ground
states \cite{magoni2023}, one for each non-symmetric triangular shape in the top
right of \figref{fig:boa}{a}. This is the so-called Jahn-Teller effect
\cite{magoni2023,jahn1937,jahn1938}. In the vicinity of the leftmost
symmetry-broken triangle, and for $\Omega=0$, the \ac{bopes} reads
\begin{equation}\label{eq:potentialEnergySurface}
    E_\mathrm{BO}(0;\bm{q})
    =
    \frac{\omega}{2x^2_0}\bm{q}^2
    +
    \frac{2\kappa}{x_0}q_\parallel
    +
    \frac{2\xi}{x^2_0}q^2_\parallel
    +
    \frac{\sqrt{2}\nu\kappa}{x^2_0}q^2_\perp
    \,,
\end{equation}%
where $q_\parallel=(q_1-q_2)\slash\sqrt{2}$ and
$q_\perp=(q_3+q_4)\slash\sqrt{2}$.

Increasing the Rabi frequency leads to a change of the molecular configuration:
the distorted triangle becomes a symmetric one [see \figref{fig:boa}{a}]. This
structural transition is reflected in the ground state energy, which is shown in
\figref{fig:boa}{b}, where we display both the full quantum mechanical result
(blue) and the \ac{boa} (red). The \ac{boa} shows a kink at the transition
point, whereas the full numerical calculation yields a smooth curve and is also
shifted. These discrepancies are quantum corrections, which are not accounted
for by the \ac{boa}. The relative shift can be computed analytically for
$\Omega=0$,
\begin{equation}\label{eq:quantumCorrections}
    E_{\mathrm{GS},3}-E^\mathrm{BO}_{\mathrm{GS},3}
    =
    \frac{\omega}{2}
    \bigg[
        \sqrt{1-\bar{\xi}}+\sqrt{1-\bar{\kappa}}-2
    \bigg]
    \,,
\end{equation}%
and the explicit calculation (see supplemental material \cite{supmat}) indeed
shows that a non-vanishing commutator between bosonic operators is its origin.
The observed smoothed transition is a consequence of the fact that in the
vicinity of the transition it is not sufficient to merely consider the minima of
the \ac{boa} to compute the ground state energy. In fact, here the formerly
neglected kinetic energy leads to tunneling between the minima and a gradual and
smooth change of the ground state energy.

\textit{Summary and conclusions.---} Optical tweezer arrays in conjunction with
Rydberg excitations offer new opportunities for engineering many-body systems
that share central features of complex molecules. Already the simple examples
discussed in this work showcase a whole host of interesting phenomena. The
quantum simulation toolbox can be further enhanced by introducing resonant (and
orientation-dependent) dipolar interactions, which yield a further mechanism ---
in addition to facilitation --- for electronic excitation hopping. This may
allow to engineer more complex molecular potential landscapes, for example,
conical intersections \cite{gambetta2021}. Further handles to enrich the quantum
simulator are state-dependent and anisotropic tweezer potentials. Ultimately,
such a quantum simulator may allow to address fundamental questions concerning
dynamical processes in extended molecules. Examples include the impact of
vibronic coupling on excitation transfer \cite{banerjee2016} in light harvesting
complexes \cite{arsenault2020}. Such systems are notoriously difficult to model
as the number of phonon degrees of freedom is too large to be exactly simulated
\cite{betti2024}. Yet, it is too small to be effectively accounted for through a
simple unstructured bath, which would enable a master equation description
\cite{eisfeld2012}.

\acknowledgments
\textit{Acknowledgments.---} We acknowledge funding from the Deutsche
Forschungsgemeinschaft within the Grant No. 452935230 and the research units
FOR5413 (Grant No. 465199066) and FOR5522 (Grant No. 499180199). This work was
also supported by the QuantERA II programme (project CoQuaDis, DFG Grant No.
532763411) that has received funding from the EU H2020 research and innovation
programme under GA No. 101017733. This work is supported by the ERC grant
OPEN-2QS (Grant No. 101164443, https://doi.org/10.3030/101164443).

\bibliography{./main.bib}

\renewcommand\thesection{S\arabic{section}}
\renewcommand\theequation{S\arabic{equation}}
\renewcommand\thefigure{S\arabic{figure}}
\setcounter{equation}{0}
\setcounter{equation}{0}
\setcounter{figure}{0}

\onecolumngrid
\clearpage
\setcounter{page}{1}

\begin{center}
    {\Large SUPPLEMENTAL MATERIAL}
\end{center}
\begin{center}
    \vspace{0.8cm}
    \Large{Rydberg atom arrays as quantum simulators for molecular dynamics}
\end{center}
\begin{center}
    Simon Euchner$^1$ and Igor Lesanovsky$^{1,2}$
\end{center}
\begin{center}
    $^1${\em Institut f\"ur Theoretische Physik and Center for Integrated
             Quantum Science and Technology, Universit\"at T\"ubingen, Auf
             der Morgenstelle 14, 72076 T\"ubingen, Germany
        }
    \\
    $^2${\em School of Physics and Astronomy and Centre for the Mathematics
             and Theoretical Physics of Quantum Non-Equilibrium Systems, The
             University of Nottingham, Nottingham, NG7 2RD, United Kingdom
        }
\end{center}

%
%
%
%
%
%
Here we present some details on results from the main text. In the first section
we define the annihilation operators in Eq.~\eqref{eq:bosonicHamiltonians} from
the main text. In the second section we derive the \acf{bt}, give the critical
vibronic coupling strengths, and illustrate how the \ac{bt} is made use of for
calculating the \acf{mgse} in the limit $\Omega=0$. In the third section we
calculate the Wigner quasiprobability distribution, which is
Eq.~\eqref{eq:wignerDistribution} from the main text. In the fourth section we
calculate the ground state atom displacements for the tetrahedron in the limit
$\Omega=0$. Finally, in the fifth section, we derive the quantum corrections,
Eq.~\eqref{eq:quantumCorrections}, from the main text. Further, the steps
involved to generate Fig.~\ref{fig:boa} from the main text are outlined.

\begin{center}
    \textbf{Parallel and perpendicular annihilation operators for the
            tetrahedron}
\end{center}

Here we give explicit expressions for the annihilation operators in
Eq.~\eqref{eq:bosonicHamiltonians} from the main text. To do so, we first define
the new vector of annihilation operators $\bm{b}=R^\mathrm{T}\bm{a}$,
\begin{equation}\label{eq:rotationTetrahedron}
    R
    =
    \mqty[ \frac{1}{\sqrt{6}} & 0 & -\frac{1}{2\sqrt{3}} &
           -\frac{1}{2\sqrt{6}} & -\frac{1}{2\sqrt{3}} & \frac{1}{\sqrt{6}} &
           -\frac{1}{2\sqrt{2}} & -\frac{1}{2\sqrt{3}} & 0
           & \frac{1}{2} & 0 & 0 \\
           -\frac{1}{3\sqrt{2}} & -\frac{1}{2\sqrt{6}} &
           -\frac{1}{6} & -\frac{\sqrt{2}}{3} & \frac{1}{6} &
           \frac{1}{6\sqrt{2}} & \frac{1}{\sqrt{6}} & -\frac{1}{3} &
           -\frac{1}{2\sqrt{3}} & 0 & \frac{1}{2} &  0 \\
           -\frac{1}{6} & \frac{1}{2\sqrt{3}} &
           -\frac{\sqrt{2}}{3} & \frac{1}{6} & -\frac{1}{3\sqrt{2}} &
           -\frac{1}{6} & \frac{1}{2\sqrt{3}} & -\frac{1}{3\sqrt{2}}
           & \frac{1}{\sqrt{6}}
           & 0 & 0 & \frac{1}{2} \\
           -\frac{1}{\sqrt{6}} & 0 & -\frac{1}{2\sqrt{3}}
           & -\frac{1}{2\sqrt{6}} & \frac{1}{2\sqrt{3}}
           & -\frac{1}{\sqrt{6}} &
           -\frac{1}{2\sqrt{2}} & \frac{1}{2\sqrt{3}} & 0
           & \frac{1}{2} & 0 & 0 \\
           -\frac{1}{3\sqrt{2}} & -\frac{1}{2\sqrt{6}} &
           \frac{1}{6} & \frac{\sqrt{2}}{3} & \frac{1}{6}
           & \frac{1}{6\sqrt{2}}
           & -\frac{1}{\sqrt{6}} & -\frac{1}{3} & \frac{1}{2\sqrt{3}}
           & 0 & \frac{1}{2} & 0 \\
           -\frac{1}{6} & \frac{1}{2\sqrt{3}}
           & \frac{\sqrt{2}}{3} & -\frac{1}{6} & -\frac{1}{3\sqrt{2}} &
           -\frac{1}{6} & -\frac{1}{2\sqrt{3}} & -\frac{1}{3\sqrt{2}} &
           -\frac{1}{\sqrt{6}} & 0 & 0 & \frac{1}{2} \\
           0 & 0 & 0 & \frac{\sqrt{6}}{4} & 0 & 0 &
           \frac{1}{2\sqrt{2}} & 0 & -\frac{1}{2} & \frac{1}{2} & 0 & 0 \\
           \frac{\sqrt{2}}{3} & -\frac{1}{2\sqrt{6}} & 0
           & 0 & -\frac{1}{3} & -\frac{5}{6\sqrt{2}} & 0 & \frac{1}{6} & 0
           & 0 & \frac{1}{2} &  0 \\
           -\frac{1}{6} & -\frac{1}{\sqrt{3}} & 0 & 0 &
           -\frac{1}{3\sqrt{2}} & \frac{1}{3} & 0 & \frac{\sqrt{2}}{3} & 0 &
           0 & 0 & \frac{1}{2} \\
           0 & 0 & \frac{1}{\sqrt{3}} &
           -\frac{1}{2\sqrt{6}} & 0 & 0 & \frac{1}{2\sqrt{2}} & 0
           & \frac{1}{2} & \frac{1}{2} & 0 & 0 \\
           0 & \frac{\sqrt{6}}{4} & 0 & 0 & 0 &
           \frac{1}{2\sqrt{2}} & 0 & \frac{1}{2} & 0 & 0 & \frac{1}{2} & 0 \\
           \frac{1}{2} & 0 & 0 & 0 & \frac{1}{\sqrt{2}} &
           0 & 0 & 0 & 0 & 0 & 0 & \frac{1}{2}
         ]
\,,
\end{equation}%
with $\bm{a}=(a^{(1)}_x,a^{(1)}_y,a^{(1)}_z,\dots,a^{(4)}_x,a^{(4)}_y,a^{(4)}_z)
^\mathrm{T}$ (transposition only acts on $\mathbb{R}^{12}$). The rotation $R$ we
obtain by diagonalizing the real-symmetric twelve-dimensional matrix
\begin{equation}\label{eq:matrixK}
    K
    =
    \mqty[  K^{(2,1)}+K^{(3,1)}+K^{(4,1)} & -K^{(2,1)}
          & -K^{(3,1)}                     & -K^{(4,1)}                     \\
           -K^{(2,1)}                     &  K^{(2,1)}+K^{(3,2)}+K^{(4,2)}
          & -K^{(3,2)}                     & -K^{(4,2)}                     \\
           -K^{(3,1)}                     & -K^{(3,2)}
          &  K^{(3,1)}+K^{(3,2)}+K^{(4,3)} & -K^{(4,3)}                     \\
           -K^{(4,1)}                     & -K^{(4,2)}
          & -K^{(4,3)}                     &  K^{(4,1)}+K^{(4,2)}+K^{(4,3)}    ]
\,,
\end{equation}%
where $K^{(k,l)}=H^{(k,l)}_\mathrm{a}+H^{(k,l)}_\mathrm{b}$ [see
Eq.~\eqref{eq:hessianAStructure} and Eq.~\eqref{eq:hessianBStructure} from the
main text]. This matrix is the Hessian of the potential $U$ that is obtained by
setting $n^{(1)}=\dots=n^{(4)}=1$ in Eq.~\eqref{eq:simulatorHamiltonian} from
the main text, i.e.,
\begin{equation}
    U(\bm{r}^{(1)}, \bm{r}^{(2)}, \bm{r}^{(3)}, \bm{r}^{(4)})
    =
      V(r^{(2,1)})
    + V(r^{(3,1)})
    + V(r^{(3,2)})
    + V(r^{(4,1)})
    + V(r^{(4,2)})
    + V(r^{(4,3)})
\,,
\end{equation}%
with $\bm{r}^{(k,l)}=\norm*{\bm{r}^{(k)}-\bm{r}^{(l)}}$. The last three columns
define $b_{10}$, $b_{11}$, and $b_{12}$. These three modes describe
center-of-mass shifts of the tetrahedron, such that they decouple due to
momentum conservation. Hence, this symmetry allows to reduce the description to
$12-3=9$ bosonic modes. Generally, $R$ does not diagonalize the interaction
Hamiltonian in Eq.~\eqref{eq:simulatorHamiltonian} from the main text. This is
because, generally, we need to employ an electronic state-dependent rotation to
define new annihilation operators such that the Hessian on each electronic
subspace is diagonal. This is best illustrated via an example. On the subspace
$\ket*{\eu\eu\ed\ed}$ the vibronic Hamiltonian is given by the expression
\begin{align}
    h_{\ket*{\eu\eu\ed\ed}}
    &=
    \omega\bm{a}^\dagger\bm{a}
    +
    \frac{x_0}{\sqrt{2}}
    [-\bm{G}^{(2,1)}, \bm{G}^{(2,1)}, \bm{0}^\mathrm{T}, \bm{0}^\mathrm{T}]
    (\bm{a}+(\bm{a}^\dagger)^\mathrm{T})
    +
    \frac{x^2_0}{4}
    (\bm{a}^\dagger+\bm{a}^\mathrm{T})
    L^{(2,1)}
    (\bm{a}+(\bm{a}^\dagger)^\mathrm{T})
\\
    &=
    \omega\bm{b}^\dagger\bm{b}
    +
    \frac{x_0}{\sqrt{2}}
    [-\bm{G}^{(2,1)}, \bm{G}^{(2,1)}, \bm{0}^\mathrm{T}, \bm{0}^\mathrm{T}]R
    (\bm{b}+(\bm{b}^\dagger)^\mathrm{T})
    +
    \frac{x^2_0}{4}
    (\bm{b}^\dagger+\bm{b}^\mathrm{T})
    R^\mathrm{T}L^{(2,1)}R
    (\bm{b}+(\bm{b}^\dagger)^\mathrm{T})
\,.
\end{align}%
Here, the vector $\bm{G}^{(2,1)}$ is defined in the main text in
Eq.~\eqref{eq:gradientStructure} and we introduced the real-symmetric matrix
\begin{equation}
    L^{(2,1)} = \mqty[  K^{(2,1)} & -K^{(2,1)} & 0_3 & 0_3 \\
                       -K^{(2,1)} &  K^{(2,1)} & 0_3 & 0_3 \\
                       0_3        &  0_3       & 0_3 & 0_3 \\
                       0_3        &  0_3       & 0_3 & 0_3    ]
\,,
\end{equation}%
where $0_3$ is the three-dimensional zero matrix. Generally,
$R^\mathrm{T}L^{(2,1)}R$ is not diagonal, such that the modes mix. In fact, it
is generally impossible to find a transformation $R$ such that
$R^\mathrm{T}L^{(k,l)}R$ is diagonal for all $k$ and $l$, because the
commutators $\comm*{L^{(k,l)}}{L^{(k',l')}}$, generally, do not vanish. In order
to cancel mixing of the modes we utilize a state-dependent second rotation
$R_{\ket*{\eu\eu\ed\ed}}$ for which
$R^\mathrm{T}_{\ket*{\eu\eu\ed\ed}}(R^\mathrm{T}L^{(2,1)}R)
R_{\ket*{\eu\eu\ed\ed}}$ is diagonal. With this second rotation we define the
state-dependent annihilation operators
$\bm{b}_{\ket*{\eu\eu\ed\ed}}=R^\mathrm{T}_{\ket*{\eu\eu\ed\ed}}\bm{b}$. Using
these yields the expression for $h_{\ket*{\eu\eu\ed\ed}}$ given in
Eq.~\eqref{eq:bosonicHamiltonians} from the main text. We apply this process for
all states $s$ representing a node in the graph depicted in
\figref{fig:arrays}{d} from the main text, such that we generally define
$\bm{b}_s=R^\mathrm{T}_s\bm{a}$. If the state $s$ hosts two Rydberg atoms, we
find that there are only three modes which are not described by a free harmonic
oscillator. These three modes are the parallel and perpendicular modes from the
main text in Eq.~\eqref{eq:bosonicHamiltonians}. Here we give the explicit
expressions for the annihilation operators associated to the subspace
$s=\ket*{\eu\eu\ed\ed}$:
\begin{equation}
    b^\parallel_{\ket*{\eu\eu\ed\ed}}
    =
    \frac{1}{\sqrt{2}} (a^{(1)}_x-a^{(2)}_x)
\,, \
    b^{\perp,1}_{\ket*{\eu\eu\ed\ed}}
    = \frac{1}{\sqrt{2}} (a^{(1)}_y-a^{(2)}_y)
\,, \
    b^{\perp,2}_{\ket*{\eu\eu\ed\ed}}
    = \frac{1}{\sqrt{2}} (a^{(1)}_z-a^{(2)}_z)
\end{equation}%
Of course this is not the only possible choice, since the rotations
$R_{\ket*{\mathrm{\eu\eu\ed\ed}}}$ and $R$ are not unique.

\begin{center}
    \textbf{Bogoliubov transformation, ground state energy, and critical
            coupling strengths}
\end{center}

Here we derive the \ac{bt} used in the main text, extract the critical coupling
strengths $\xi_\mathrm{c}$ and $\kappa_\mathrm{c}$, and show how the ground
state energy of the arrays in \figref{fig:arrays}{a-c} from the main text can be
calculated in the limit $\Omega=0$.

We start with the \ac{bt}. To this end we consider the generic bosonic
Hamiltonian
\begin{equation}\label{eq:genericBosonicHamiltonian}
    h = \omega b^\dagger b + \lambda (b^\dagger+b) + \gamma (b^\dagger+b)^2
\,,
\end{equation}%
with $\omega>0$ and $\lambda,\gamma\in\mathbb{R}$. Appropriate substitutions of
the parameters $\lambda$ and $\gamma$ yield the perturbed harmonic oscillators
in Eq.~\eqref{eq:dumbbellHamiltonian} and Eq.~\eqref{eq:bosonicHamiltonians}
from the main text. Our goal is to find a \ac{bt} $b_\mathrm{BT}=ub+vb^\dagger$,
with $u,v\in\mathbb{C}$ such that the quadratic off-diagonal terms vanish.
Without loss of generality we choose $u\geq{0}$. Note that $b_\mathrm{BT}$ obeys
bosonic algebra if and only if $\abs*{u}^2-\abs*{v}^2=1$, such that $u=0$ is not
a solution. Therefore, we do not loose solutions by assuming $u>0$. To obtain
equations which fix $u$ and $v$ we first substitute the inverse transformation
$b=u^*b^\dagger_\mathrm{BT}-vb_\mathrm{BT}$ into
Eq.~\eqref{eq:genericBosonicHamiltonian}. The result is
\begin{multline}\label{eq:transformedGenericBosonicHamiltonian}
    h = \big(
            (\omega+2\gamma)(\abs*{u}^2+\abs*{v}^2)-2\gamma(uv^*+u^*v)
        \big)
        b^\dagger_\mathrm{BT} b_\mathrm{BT}
        +
        \lambda ((u-v)b^\dagger_\mathrm{BT}+(u^*-v^*)b_\mathrm{BT})
\\
        +
        (f(u,v)b^{\dagger 2}_\mathrm{BT}+f^*(u,v)b^2_\mathrm{BT})
        +
        (\omega+2\gamma)\abs*{v}^2+\gamma(1-uv^*-u^*v)
\,.
\end{multline}%
Here we defined the function $f(u,v)=\gamma(u^2+v^2)-(\omega+2\gamma)uv$. To
achieve our goal we choose $u$ and $v$ such that $f(u,v)=0$. Since $u>0$, the
ratio $w=v\slash{u}$ is well defined and we find that $f(u,v)=0$ is equivalent
to the condition
\begin{equation}
    \gamma w^2 - (\omega+2\gamma)w + \gamma = 0
\,.
\end{equation}%
For $\gamma=0$, Eq.~\eqref{eq:genericBosonicHamiltonian} does not host terms
like $b^2$ and $(b^\dagger)^2$, such that the \ac{bt} is trivial. Let now
$\gamma\neq{0}$. Solving for $w$ yields
\begin{equation}
    w
    =
    \begin{cases}
        1+\phi\pm\sqrt{(1+\phi)^2-1}\,, \ &(1+\phi)^2 > 1
\\
        1+\phi\pm\ii\sqrt{1-(1+\phi)^2}\,, \ &(1+\phi)^2 \leq 1
    \end{cases}
\,, \
    \phi = \frac{\omega}{2\gamma}
\,.
\end{equation}%
To obtain the \ac{bt} we further enforce the condition
$\abs*{u}^2-\abs*{v}^2=1$, which turns out to be impossible if
$(1+\phi)^2\leq{1}$, since $\abs*{w}=1$. In case $(1+\phi)^2>1$ we always find a
solution that obeys $\abs*{w}<1$. The condition for solutions to exist can thus
be expressed as $\phi\in(-\infty,-2)\cup(0,\infty)$. Therefore, we close (again
allowing the case $\gamma=0$) that the \ac{bt} exists if and only if
$2\gamma\slash\omega\in(-1\slash{2},\infty)$. In the regime were a solution for
$u$ and $v$ exists, we obtain
\begin{equation}\label{eq:uvBogoliubovTransformation}
    u = \frac{1}{\sqrt{1-w^2}}
\,, \
    v = \frac{w}{\sqrt{1-w^2}}
\,,
\end{equation}%
where
\begin{equation}\label{eq:generalSolutionW}
    w = 1+\phi-\frac{\phi}{\abs{\phi}}\sqrt{(1+\phi)^2-1}
\,, \
    \phi = \frac{\omega}{2\gamma}
\,,
\end{equation}%
which is Eq.~\eqref{eq:solutionForW} from the main text.

The condition that determines if the \ac{bt} exists, yields the critical
parameter $\gamma_\mathrm{c}=-\omega\slash{4}$. For $\gamma=\xi$ [cf.
Eq.~\eqref{eq:dumbbellHamiltonian} from the main text] we have
$\xi_\mathrm{c}=-\omega\slash{4}$. In the case $\lambda=0$ and
$\gamma=\nu\kappa\slash\sqrt{2}$ we recover the Hamiltonian associated to the
annihilation operator $b^{\perp,1}_s$ [cf. Eq.~\eqref{eq:bosonicHamiltonians}
from the main text]. In this case $\gamma_\mathrm{c}=-\omega\slash{4}$ results
in $\nu\kappa_\mathrm{c}\slash\sqrt{2}=-\omega\slash{4}$, i.e., the critical
coupling strength $\kappa_\mathrm{c}=-\omega\slash(2\sqrt{2}\nu)$.

With the \ac{bt} at hand it is now straight forward to calculate the \ac{mgse}
in the limit $\Omega=0$ for the arrays in \figref{fig:arrays}{a-c} from the main
text. As an example we illustrate this for the dumbbell. In the limit $\Omega=0$
there are two possibilities for the ground state: either
$\ket*{+}\otimes\ket*{\mathrm{vac}}_b$, where $\ket*{\mathrm{vac}}_b$ is the
vacuum of the annihilation operator $b$, or
$\ket*{\eu\eu}\otimes\ket*{\mathrm{gs}}$, with the ground state
$\ket*{\mathrm{gs}}$ of the perturbed harmonic oscillator
\begin{equation}\label{eq:dumbellTransformedHamil}
    H = \omega b^\dagger b + \sqrt{2}\kappa(b+b^\dagger) + \xi(b+b^\dagger)^2
\,.
\end{equation}%
If the ground state is $\ket*{+}\otimes\ket*{\mathrm{vac}}_b$, the ground state
energy is zero. Therefore, the ground state is given by
$\ket*{+}\otimes\ket*{\mathrm{vac}}_b$ only if the ground state energy of $H$ is
larger than zero. If the ground state energy of $H$ is zero, the ground state is
twofold-degenerate. Combining all of this results in the expression
$E_{\mathrm{GS},2}=\mathrm{min}\{\omega\varepsilon_s,0\}$ for the \ac{mgse},
where $\omega\varepsilon_2$ is the ground state energy of $H$. To determine
$\varepsilon_2$ we first apply the \ac{bt} to $H$, which leads to
\begin{equation}
    H
    =
    \omega\sqrt{1-\bar{\xi}}b^\dagger_\mathrm{BT}b_\mathrm{BT}
    +
    \frac{\sqrt{2}\kappa}{(1-\bar{\xi})^{1\slash{4}}}
    (b_\mathrm{BT}+b^\dagger_\mathrm{BT})
    +
    \frac{\omega}{2}\sqrt{1-\bar{\xi}}-\frac{\omega}{2}
\,, \
    \bar{\xi} = \frac{\xi}{\xi_\mathrm{c}}
\,.
\end{equation}%
Next, we transform into a displaced frame of reference using the displacement
operator
\begin{equation}\label{eq:diplacementOperator}
    D(b_\mathrm{BT},\mu) = \exp(\mu b^\dagger_\mathrm{BT} - \mu^* b_\mathrm{BT})
\,, \
    \mu = -\frac{\sqrt{2}\kappa}{\omega}\frac{1}{(1-\bar{\xi})^{3\slash{4}}}
\,.
\end{equation}%
In the displaced frame we find
\begin{equation}
    D^\dagger(b_\mathrm{BT},\mu)
    H
    D(b_\mathrm{BT},\mu)
    =
    \omega \sqrt{1-\bar{\xi}}b^\dagger_\mathrm{BT} b_\mathrm{BT}
    -\frac{2\kappa^2}{\omega}\frac{1}{1-\bar{\xi}}
    +
    \frac{\omega}{2}\sqrt{1-\bar{\xi}}-\frac{\omega}{2}
\,.
\end{equation}%
Clearly, the ground state of the displaced Hamiltonian is the vacuum of
$b_\mathrm{BT}$, such that we identify $\omega\varepsilon_2$ to be the additive
constant to the free harmonic oscillator, which is
Eq.~\eqref{eq:groundStateEnergyDumbbell} from the main text. Analogously, we
find $\varepsilon_4$ in Eq.~\eqref{eq:groundStateEnergyTetrahedron} from the
main text.

\begin{center}
    \textbf{Wigner quasiprobability distribution}
\end{center}

In this section we derive the Wigner quasiprobability distribution that is given
in Eq.~\eqref{eq:wignerDistribution} from the main text. This means that we
calculate it for the pure state $\rho_\mathrm{gs}=\dyad*{\mathrm{gs}}$ and the
annihilation operator $c=b_{\ket*{\eu\eu\ed\ed}}^{\perp,1}$. Here,
$\ket*{\mathrm{gs}}$ is the ground state of the Hamiltonian [see
Eq.~\eqref{eq:bosonicHamiltonians} from the main text] associated to the
annihilation operator $c$, i.e.,
\begin{equation}
    H = \omega c^\dagger c + \frac{\nu\kappa}{\sqrt{2}}(c+c^\dagger)^2
\,.
\end{equation}%
Note that the state $\ket*{\mathrm{GS}}$ from the main text is not equal to
$\ket*{\mathrm{gs}}$, because $\ket*{\mathrm{GS}}$ is the ground state of
$h_{\ket*{\eu\eu\ed\ed}}$, not just the ground state of $H$. The first step now
is to calculate $\ket*{\mathrm{gs}}$. To this end we rewrite it in terms of the
new annihilation operator $c_\mathrm{BT}=(c+wc^\dagger)\slash\sqrt{1-w^2}$,
where $w$ is given by Eq.~\eqref{eq:generalSolutionW} with the substitution
$\gamma=\nu\kappa\slash\sqrt{2}$. Making use of
Eq.~\eqref{eq:transformedGenericBosonicHamiltonian}, we arrive at
\begin{equation}
    H
    =
    \omega\sqrt{1-\bar{\kappa}}
    c^\dagger_\mathrm{BT}c_\mathrm{BT}
    +
    \frac{\omega}{2}\sqrt{1-\bar{\kappa}}-\frac{\omega}{2}
\,, \
    \bar{\kappa} = \frac{\kappa}{\kappa_\mathrm{c}}
\,.
\end{equation}%
From this expression we close that the ground state is given by the vacuum of
$c_\mathrm{BT}$, i.e. $\ket*{\mathrm{gs}}=\ket*{\mathrm{vac}}_{c_\mathrm{BT}}$.
To relate this state to Fock states associated to the annihilation operator $c$,
we use that the \ac{bt} acts on the Hilbert space in the form of a squeezing
operator:
\begin{equation}\label{eq:phaseSpaceBT}
    c_\mathrm{BT} = S^\dagger(c,\sigma) c S(c,\sigma)
\,, \
    S(c,\sigma) = \exp(\frac{\sigma^*}{2}c^2-\frac{\sigma}{2}(c^\dagger)^2)
\end{equation}%
In our case, $\sigma=-\mathrm{artanh}(w)$, with $w$ given by
Eq.~\eqref{eq:generalSolutionW} for $\gamma=\nu\kappa\slash\sqrt{2}$. Note that
this is well defined because $\abs*{w}<1$ if the \ac{bt} exists. Since by
definition the ground state is annihilated by $c$,
\begin{equation}\label{eq:vacuumBT}
    0
    =
    S^\dagger(c,\sigma)c\ket*{\mathrm{vac}}_c
    =
    S^\dagger(c,\sigma)S(c,\sigma)
    c_\mathrm{BT}S^\dagger(c,\sigma)\ket*{\mathrm{vac}}_c
    =
    c_\mathrm{BT}S^\dagger(c,\sigma)\ket*{\mathrm{vac}}_c
\,.
\end{equation}%
Up to a phase, this yields the relation
$\ket*{\mathrm{vac}}_{c_\mathrm{BT}}=S^\dagger(c,\sigma)\ket*{\mathrm{vac}}_c$.
Therefore, in the ground state $\rho_\mathrm{gs}$, we find that the expectation
value of the displacement operator [defined in
Eq.~\eqref{eq:diplacementOperator}]
\begin{equation}\label{eq:displacer}
    D(c,\beta)=\exp(\beta{c^\dagger}-\beta^*c)
\end{equation}%
acquires the form
\begin{equation}
    f(\beta)
    =
    \mathrm{tr}(\rho_\mathrm{gs}D(c,\beta))
    =
    \exp(-\frac{1}{2}\frac{\abs*{\beta+w\beta^*}}{1-w^2})
\,,
\end{equation}%
with $\beta\in\mathbb{C}$. Finally, to obtain the Wigner quasiprobability
distribution $W$, we essentially take the complex Fourier transformation of $f$:
\begin{equation}
    W(\alpha)
    =
    \frac{1}{\pii^2}
    \int_\mathbb{C}\dd{\beta}
    f(\beta)
    \ee^{\beta^*\alpha-\alpha^*\beta}
\end{equation}%
The integration results in Eq.~\eqref{eq:wignerDistribution} from the main text.

\begin{center}
    \textbf{Ground state displacements at vanishing Rabi frequency}
\end{center}

In this section we justify the claim that in the state
$\ket*{\eu\eu\ed\ed}\otimes\ket*{\mathrm{GS}}$, that is used to calculate the
Wigner quasiprobability distribution, the atoms are displaced less than $0.4x_0$
from their equilibrium position, for the vibronic couplings given in the caption
of \figref{fig:instabilities}{c,d} and $\kappa=0.99\kappa_\mathrm{c}$. To this
end, we derive an analytical expression for the ground state displacements in
this state. The first step is to obtain $\ket*{\mathrm{GS}}$, which is the
ground state of the Hamiltonian [see Eq.~\eqref{eq:bosonicHamiltonians} in the
main text]
\begin{multline}
    h_{\ket*{\eu\eu\ed\ed}}
    =
    \omega\bm{b}^\dagger_{\ket*{\eu\eu\ed\ed}}\bm{b}_{\ket*{\eu\eu\ed\ed}}
    +
    \sqrt{2}\kappa
    (   b^\parallel_{\ket*{\eu\eu\ed\ed}}
      + (b^\parallel_{\ket*{\eu\eu\ed\ed}})^\dagger )
    +
    \xi
    (   b^\parallel_{\ket*{\eu\eu\ed\ed}}
      + (b^\parallel_{\ket*{\eu\eu\ed\ed}})^\dagger )^2
\\
    +
    \frac{\nu\kappa}{\sqrt{2}}
    (   b^{\perp,1}_{\ket*{\eu\eu\ed\ed}}
      + (b^{\perp,1}_{\ket*{\eu\eu\ed\ed}})^\dagger )^2
    +
    \frac{\nu\kappa}{\sqrt{2}}
    (   b^{\perp,2}_{\ket*{\eu\eu\ed\ed}}
      + (b^{\perp,2}_{\ket*{\eu\eu\ed\ed}})^\dagger )^2
\,.
\end{multline}%
The ground state of the modes which are described by free harmonic oscillators
is the vacuum, such that they do not yield finite displacements. Next, we
observe that the ground state associated to the perpendicular modes is a
squeezed vacuum, because there are no linear contributions in
$b^{\perp,j}_{\ket*{\eu\eu\ed\ed}}$. Therefore,
$\mel*{\mathrm{GS}}{(b^{\perp,j}_s+(b^{\perp,j}_s)^\dagger)}{\mathrm{GS}}=0$.
From this we close that only $b^\parallel_{\ket*{\eu\eu\ed\ed}}$ yields finite
displacements. Calculating them amounts to evaluating the expectation value
\begin{equation}\label{eq:GSdisplParallelMode}
    \frac{\delta}{x_0}
    =
    \frac{1}{\sqrt{2}}
    \mel*{\mathrm{gs}}
         {   b^\parallel_{\ket*{\eu\eu\ed\ed}}
           + (b^\parallel_{\ket*{\eu\eu\ed\ed}})^\dagger }
         {\mathrm{gs}}
\,,
\end{equation}%
where $\ket*{\mathrm{gs}}$ is the ground state of the single-mode Hamiltonian
\begin{equation}
    H
    =
    \omega c^\dagger c
    +
    \sqrt{2}\kappa(c+c^\dagger)
    +
    \xi(c+c^\dagger)^2
\,, \
    c \equiv b^\dagger_{\ket*{\eu\eu\ed\ed}}
\,.
\end{equation}%
We note that this is the same expression as in
Eq.~\eqref{eq:dumbellTransformedHamil}. Hence, we find
\begin{equation}
    D^\dagger(c_\mathrm{BT},\mu)
    H
    D(c_\mathrm{BT},\mu)
    =
    \omega \sqrt{1-\bar{\xi}}c^\dagger_\mathrm{BT} c_\mathrm{BT}
    -\frac{2\kappa^2}{\omega}\frac{1}{1-\bar{\xi}}
    +
    \frac{\omega}{2}\sqrt{1-\bar{\xi}}-\frac{\omega}{2}
\,,
\end{equation}%
with the displacement operator $D(c_\mathrm{BT},\mu)$ defined in
Eq.~\eqref{eq:diplacementOperator}. The ground state in this frame is the
vacuum, $\ket*{\mathrm{vac}}_{c_\mathrm{BT}}$, of the annihilation operator
$c_\mathrm{BT}$. Transforming back into the non-displaced frame yields the
ground state
\begin{equation}
    \ket*{\mathrm{gs}}
    =
    D(c_\mathrm{BT},\mu)S^\dagger(c,\sigma)\ket*{\mathrm{vac}}_c
    =
    D(S^\dagger(c,\sigma) c S(c,\sigma),\mu)
    S^\dagger(c,\sigma)\ket*{\mathrm{vac}}_c
    =
    S^\dagger(c,\sigma)D(c,\mu)\ket*{\mathrm{vac}}_c
\,,
\end{equation}%
with the squeezing operator [cf. Eq.~\eqref{eq:phaseSpaceBT}] $S(c,\sigma)$
associated to $c$ and $\sigma=-\mathrm{artanh}(w)$, where $w$ is given by
Eq.~\eqref{eq:generalSolutionW} after the substitution $\gamma=\xi$. Employing
this expression, we calculate
\begin{equation}
    \frac{\delta}{x_0}
    =
    \frac{1}{\sqrt{2}}
    \bra*{\mathrm{vac}}_c
    D^\dagger(c,\mu)S(c,\sigma)
    (c+c^\dagger)
    S^\dagger(c,\sigma)D(c,\mu)
    \ket*{\mathrm{vac}}_c
    =
    \frac{1}{\sqrt{2}}
    \bra*{\mu}_cS(c,\sigma)
    (c+c^\dagger)
    S^\dagger(c,\sigma)\ket*{\mu}_c
\,,
\end{equation}%
where $\ket*{\mu}_c$ denotes a coherent state of amplitude $\mu$ with respect to
$c$. Next, we combine $S^\dagger(\sigma,b)=S(-\sigma,b)$,
Eq.~\eqref{eq:phaseSpaceBT}, and Eq.~\eqref{eq:uvBogoliubovTransformation} to
arrive at
\begin{equation}
    \frac{\delta}{x_0}
    =
    \frac{1}{\sqrt{2}}
    \bra*{\mathrm{gs}}
    \big(c+c^\dagger\big)
    \ket*{\mathrm{gs}}
    =
    \frac{1}{\sqrt{2}}
    \sqrt{\frac{1-w}{1+w}}
    \bra*{\mu}_c
    (c+c^\dagger)
    \ket*{\mu}_c
    =
    \frac{1}{\sqrt{2}}
    \sqrt{\frac{1-w}{1+w}}
    (\mu+\mu^*)
\,.
\end{equation}%
Finally, plugging in the explicit expressions for $\mu$ [see
Eq.~\eqref{eq:diplacementOperator}] and $w$ [see Eq.~\eqref{eq:generalSolutionW}
with $\gamma=\xi$] results in
\begin{equation}
    \frac{\delta}{x_0}
    =
    -\frac{2\kappa}{\omega}\frac{1}{1-\bar{\xi}}
\,.
\end{equation}%
Note that because of symmetry, on average, each of the two Rydberg atoms
associated to the state $\ket*{\eu\eu\ed\ed}$ is displaced
$\abs*{\delta}\slash{2}$ from its equilibrium position. For the vibronic
couplings $\kappa=0.99\kappa_\mathrm{c}$, $\xi=2.0\omega$, and $\nu=0.10$ [see
caption of \figref{fig:instabilities}{c,d}] we obtain
$\abs*{\delta}\slash{2}\approx{0.39}x_0<0.40x_0$, which is the value given in
the main text.

\begin{center}
    \textbf{Born-Oppenheimer approximation for the triangle}
\end{center}

This final section's purpose is twofold: first, we derive
Eq.~\eqref{eq:quantumCorrections} from the main text, showing that, indeed, this
contribution is the consequence of the non-vanishing bosonic commutators, and
secondly, we want to outline the numerical procedure which we apply to produce
Fig.~\ref{fig:boa} from the main text.

To obtain Eq.~\eqref{eq:quantumCorrections} from the main text, we need the
exact expression, $E_{\mathrm{GS},3}$, for the \ac{mgse} as well as the result
in the \acf{boa}, $E^\mathrm{BO}_{\mathrm{GS},3}$. We first calculate the exact
result. To this end, analogously to the tetrahedron, we define the annihilation
operators $\bm{b}=R^\mathrm{T}\bm{a}$,
\def\ea{\frac{1}{2}}
\def\eb{\frac{1}{2\sqrt{3}}}
\def\ec{\frac{1}{\sqrt{3}}}
\begin{equation}
    R = \mqty[ \ea & -\ea &  \eb & -\eb & \ec &   0 \\
              -\eb & -\eb & -\ea & -\ea & 0   & \ec \\
              -\ea &  \ea &  \eb & -\eb & \ec &   0 \\
              -\eb & -\eb &  \ea &  \ea & 0   & \ec \\
               0   &  0   & -\ec &  \ec & \ec & 0   \\
               \ec &  \ec &  0   &  0   & 0   & \ec    ]
\,.
\end{equation}%
The rotation $R$ diagonalizes the six-dimensional matrix
\begin{equation}
    K = \mqty[    K^{(2,1)}+K^{(3,1)} & -K^{(2,1)}
               & -K^{(3,1)}           \\
                 -K^{(2,1)} &  K^{(2,1)}+K^{(3,2)}
               & -K^{(3,2)}           \\
                 -K^{(3,1)}           & -K^{(3,2)}
               &  K^{(3,1)}+K^{(3,2)}    ]
\,.
\end{equation}%
Here we defined the vector of annihilation operators
$\bm{a}=(a^{(1)}_x,a^{(1)}_y,a^{(2)}_x,a^{(2)}_y,a^{(3)}_x,a^{(3)}_y)
^\mathrm{T}$, where transposition acts solely on $\mathbb{R}^6$. The modes
associated to $b_5$ and $b_6$ decouple because of momentum conservation, such
that only four modes remain as a consequence of this symmetry. To obtain the
\ac{mgse} in the limit $\Omega=0$ we note that the Hamiltonian for the triangle
has the same form as Eq.~\eqref{eq:tetrahedronHamiltonian} from the main text.
The ground state is three-fold degenerate. Depending on the vibronic couplings
$\kappa$, $\xi$, and $\nu$, the eigenspace associated to the \ac{mgse} is either
spanned by
\begin{equation}\label{eq:triangleSingleExcitedGSs}
    \ket*{\eu\ed\ed}\otimes\ket*{\mathrm{vac}}_{\bm{b}}
\,, \
    \ket*{\ed\eu\ed}\otimes\ket*{\mathrm{vac}}_{\bm{b}}
\,, \
    \ket*{\ed\ed\eu}\otimes\ket*{\mathrm{vac}}_{\bm{b}}
\,.
\end{equation}%
or
\begin{equation}\label{eq:triangleDoubleExcitedGSs}
    \ket*{\eu\eu\ed}\otimes\ket*{\mathrm{gs}}_{\eu\eu\ed}
\,, \
    \ket*{\ed\eu\eu}\otimes\ket*{\mathrm{gs}}_{\ed\eu\eu}
\,, \
    \ket*{\eu\ed\eu}\otimes\ket*{\mathrm{gs}}_{\eu\ed\eu}
\,,
\end{equation}%
where $\ket*{\mathrm{vac}}_{\bm{b}}$ is the multimode vacuum of the annihilation
operators $\bm{b}$ and $\ket*{\mathrm{gs}}_s$ is the ground state of the
Hamiltonian $h_s$ [cf. Eq.~\eqref{eq:tetrahedronHamiltonian} from the main
text]. As an example, for $s=\ket*{\eu\eu\ed}$, the vibronic Hamiltonian is
\begin{equation}\label{eq:triangelUUDSubsoace}
    h_{\ket*{\eu\eu\ed}}
    =
    \omega\bm{b}^\dagger\bm{b}
    +
    \sqrt{2}\kappa
    (b^\parallel_{\ket*{\eu\eu\ed}}+(b^\parallel_{\ket*{\eu\eu\ed}})^\dagger)
    +
    \xi
    (b^\parallel_{\ket*{\eu\eu\ed}}+(b^\parallel_{\ket*{\eu\eu\ed}})^\dagger)^2
    +
    \frac{\nu\kappa}{\sqrt{2}}
    (b^\perp_{\ket*{\eu\eu\ed}}+(b^\perp_{\ket*{\eu\eu\ed}})^\dagger)^2
\,,
\end{equation}%
where
\begin{equation}
    b^\parallel_{\ket*{\eu\eu\ed}}
    =
    \frac{1}{\sqrt{2}}(b_1-b_2)
    =
    \frac{1}{\sqrt{2}}(a^{(1)}_x-a^{(2)}_x)
\,, \
    b^\perp_{\ket*{\eu\eu\ed}}
    =
    \frac{1}{\sqrt{2}}(b_3+b_4)
    =
    -\frac{1}{\sqrt{2}}(a^{(1)}_y-a^{(2)}_y)
\,.
\end{equation}%
If the ground state is a linear combination of the states in
Eq.~\eqref{eq:triangleSingleExcitedGSs}, the \ac{mgse} is zero. Hence, we close
that the ground state is a linear combination of the states in
Eq.~\eqref{eq:triangleDoubleExcitedGSs} only if the ground state energy,
$\omega\varepsilon_3$, of $h_{\ket*{\eu\eu\ed\ed}}$ is less than zero. We
summarize this discussion in the expression
$E_{\mathrm{GS},3}=\mathrm{min}\{\omega\varepsilon_3,0\}$ for the \ac{mgse}.
What is left to do now is to find an expression for $\omega\varepsilon_3$. To
this end we diagonalize Eq.~\eqref{eq:triangelUUDSubsoace}, which we achieve by
employing the \ac{bt} defined by Eq.~\eqref{eq:uvBogoliubovTransformation} and
Eq.~\eqref{eq:generalSolutionW}. We apply this transformation to the
Hamiltonians associated to the parallel, $b^\parallel_{\ket*{\eu\eu\ed}}$, and
perpendicular, $b^\perp_{\ket*{\eu\eu\ed}}$, annihilation operator, separately.
Both of the resulting singe-mode Hamiltonians are analogous to
Eq.~\eqref{eq:dumbellTransformedHamil}, such that we directly obtain
\begin{equation}\label{eq:groundStateEnergyTriangle}
    \varepsilon_3
    =
    -\frac{2\kappa^2}{\omega^2}\frac{1}{1-\bar{\xi}}
    +
    \frac{1}{2}\sqrt{1-\bar{\xi}}
    +
    \frac{1}{2}\sqrt{1-\bar{\kappa}}-1
\,.
\end{equation}%
Apart from the first one, all of these terms can be traced back to
\begin{equation}
    (\omega+2\gamma)\abs*{v}^2+\gamma(1-uv^*-u^*v)
\end{equation}%
in Eq.~\eqref{eq:transformedGenericBosonicHamiltonian}. Importantly, this
contribution results from the non-vanishing commutators of the bosonic algebra.
The next step is to calculate the \ac{mgse} in the \ac{boa}. To this end we
express the full Hamiltonian in the quadratures
$q_i=x_0(b_i+b^\dagger_i)\slash\sqrt{2}$, $i=1,2,3,4$ ($b_5$ and $b_6$ are
decoupled because of momentum conservation), and their conjugated momenta $p_i$.
We collect the positions and momenta in the vectors $\bm{q}$ and $\bm{p}$.
Expressed in them, the full Hamiltonian acquires the form
\begin{align}\label{eq:BOhamiltonianTriangle}
    \begin{split}
    H_3
    &=
    \Omega\sum_{s s'}B_{ss'}\dyad{s}{s'}
    +
    \frac{\omega}{2}\bigg[\frac{\bm{q}^2}{x_0^2}+x_0^2\bm{p}^2\bigg]
    \\ &\phantom{=}+
    \dyad*{\eu\eu\ed}
    \bigg[
        \frac{\sqrt{2}\kappa}{x_0} (q_1-q_2)
        +
        \frac{\xi}{x_0^2} (q_1-q_2)^2
        +
        \frac{\nu\kappa}{\sqrt{2}x_0^2} (q_3+q_4)^2
    \bigg]
    \\ &\phantom{=}+
    \dyad*{\ed\eu\eu}
    \bigg[
        \frac{\sqrt{2}\kappa}{x_0}
        \bigg(q_1+\frac{1}{2}q_2-\frac{\sqrt{3}}{2}q_3\bigg)
        +
        \frac{\xi}{x_0^2}\bigg(q_1+\frac{1}{2}q_2-\frac{\sqrt{3}}{2}q_3\bigg)^2
        +
        \frac{\nu\kappa}{\sqrt{2}x_0^2}
        \bigg(\frac{\sqrt{3}}{2}q_2+\frac{1}{2}q_3-q_4\bigg)^2
    \bigg]
    \\ &\phantom{=}+
    \dyad*{\eu\ed\eu}
    \bigg[
        \frac{\sqrt{2}\kappa}{x_0}
        \bigg(q_1+\frac{1}{2}q_2+\frac{\sqrt{3}}{2}q_3\bigg)
        +
        \frac{\xi}{x_0^2}
        \bigg(q_1+\frac{1}{2}q_2+\frac{\sqrt{3}}{2}q_3\bigg)^2
        +
        \frac{\nu\kappa}{\sqrt{2}x_0^2}
        \bigg(-\frac{\sqrt{3}}{2}q_2+\frac{1}{2}q_3-q_4\bigg)^2
    \bigg]
\,.
    \end{split}
\end{align}%
Applying the \ac{boa}, which amounts to setting $\bm{p}=\bm{0}$, allows to treat
$\bm{q}$ classically. Next, we define the ground state potential energy surface
as the map $E_\mathrm{BO}$, which maps a position $\bm{q}$ to the smallest
eigenvalue of Eq.~\eqref{eq:BOhamiltonianTriangle} for $\bm{p}=\bm{0}$. The
\ac{mgse} is then approximated by the minimum value of this surface. In the
limit $\Omega=0$ we can give an analytical expression for $E_\mathrm{BO}$, which
is Eq.~\eqref{eq:potentialEnergySurface} from the main text. Since this is a
simple parabola we obtain for the minimum
\begin{equation}\label{eq:energyTriangleBoa}
    E^\mathrm{BO}_{\mathrm{GS},3}
    =
    -\frac{2\kappa^2}{\omega}\frac{1}{1-\bar{\xi}}
\,.
\end{equation}%
With this expression we recover Eq.~\eqref{eq:quantumCorrections} from the main
text by subtracting Eq.~\eqref{eq:energyTriangleBoa} from
Eq.~\eqref{eq:groundStateEnergyTriangle}. Note that the vibronic couplings given
in the caption of \figref{fig:boa}{a} from the main text yield
$E_{\mathrm{GS},3}=\omega\varepsilon_3<0$.

Finally, we outline how Fig.~\ref{fig:boa} from the main text is generated.
First, we fix all vibronic couplings to the values presented in the caption of
the figure. With these constants fixed, $E_\mathrm{BO}$ is now a function of
$\Omega$ and $\bm{q}$. We then use a standard minimization algorithm to obtain
the (approximate) \ac{mgse} for a range of $\Omega$. The resulting (red) curve
is shown in \figref{fig:boa}{b} from the main text. To obtain the three curves
in \figref{fig:boa}{a} from the main text, we first note that in the
symmetry-broken regime $E_\mathrm{BO}$ hosts three degenerate global minima,
corresponding to the symmetry-broken triangular configurations depicted in the
top right (see also \cite{magoni2023}). To show the transition we plot
$E_\mathrm{BO}$ in the vicinity of the global minimum which is associated with
the leftmost symmetry-broken triangle. At this specific minimum $q_3=q_4=0$. In
the remaining two-dimensional space we then evaluate $E_\mathrm{BO}$ along the
curve $q_{1,\mathrm{min}}(q_2)$. This curve maps $q_2$ to the particular $q_1$
for which $E_\mathrm{BO}(\Omega;q_1,q_2,q_3=0,q_4=0)$ is minimal. This we do for
three different $\Omega$ [yellow markers in \figref{fig:boa}{b}], which results
in the solid, the dashed, and the dash-dotted line in \figref{fig:boa}{a}.


\end{document}